# Supporting and Controlling Complex Concurrency in Fault-Tolerant Distributed Systems

*J. Xu, B. Randell, A. Romanovsky, R.J. Stroud and A.F. Zorzo*

Department of Computing Science
University of Newcastle upon Tyne, Newcastle upon Tyne, UK

**Abstract**

*Distributed computing often gives rise to complex concurrent and interacting activities. In some cases several concurrent activities may be working together, i.e. cooperating, to solve a given problem; in other cases the activities may be independent but needing to share common system resources for which they must compete. Many difficulties and limitations occur in the widely advocated objects and (trans)actions model when it is supposed to support cooperating activities. We have introduced previously the concept of coordinated atomic (CA) actions [Xu et al. 1995]; this paper analyzes and examines the derived objects and CA actions model for constructing fault-tolerant distributed systems and providing unified support for both cooperative and competitive concurrency. Our investigation reveals and clarifies several significant problems that have not previously been studied extensively, including the problem of ensuring consistent access to shared objects from a joint action as opposed to a set of independent actions. Conceptual and implementation-related solutions are proposed and illustrated.*

## 1: Introduction

A widely-used computational model for introducing fault tolerance — particularly in distributed object-oriented systems — is based on the use of *atomic (trans)actions* [Gray 1978][Gray & Reuter 1993][Shrivastava & McCue 1994] that control operations on shared objects with the properties of failure atomicity, serializability and permanence of effect. This model attempts to hide the effects of failures and concurrent processing from the application programmer, and is very effective for certain applications in which concurrent activities are relatively independent, just needing to share some common objects for which they have to *compete*. Practical examples include some types of electronic funds transfer, office information, and airline reservation systems. However, through the widespread use of the objects and actions model, limitations and difficulties with it are becoming evident. One of its severe limitations is that the model does not provide appropriate support for cooperation between concurrent activities.

In many real-world environments, there are specific needs for cooperation and coordination among activities. For example, several concurrent activities may need to work together, *cooperate*, to solve a given problem. Since such cooperative concurrency constitutes explicit interactions among activities, some way (in principle more sophisticated than the traditional transaction concept) of enclosing interactions will be desirable in order to control the overall system complexity and hence to facilitate error recovery. The best-known approach to structuring (cooperatively) concurrent systems is the *conversation* concept [Randell 1975]. Unfortunately, conversations are mainly directed to process-oriented systems such as process control or real-time control applications, and concentrate on the issues of factoring and organizing the interactions between processes, and do not address the problem of consistency of shared objects accessed by these processes.

Researchers are now working hard to merge various existing models and approaches into a new unified concept able to handle a variety of applications including databases, process control, and computer-supported cooperative work. To the best of our knowledge, there are at least two independent strands of development relevant to such research efforts. In the area of transactions and databases, generalized transaction models have been introduced to overcome the limitations of the traditional transaction concept and to allow certain cooperation between activities (or between nested transactions) [Elmagarmid 1993][Taylor *et al.* 1994]. Most generalized models use the same structure as the nested transaction model [Moss 1981], but concurrent nested transactions may be allowed to cooperate by relaxing strict isolation — uncommitted results can be shared among nested transactions — or by enforcing user-defined order — execution of nested transactions or certain operations is specially ordered. Two common disadvantages of these models are 1) loss of the atomicity property of nested transactions and 2) limited support for cooperation. (Section three will contain a more detailed discussion.)

The concept of *coordinated atomic actions* (or CA actions), recently developed at Newcastle University and especially for object-oriented system design [Xu *et al.* 1995], is the first attempt to integrate transactions and conversations into a single conceptual framework for coping with different kinds of concurrency. CA actions provide a control abstraction that encloses interactions among multiple cooperative activities; there are two major distinctions between the CA action concept and those generalized transaction models — 1) the atomicity property that holds at any level of nesting, and 2) full flexibility in defining application-specific interactions and coordinations. However, the CA action concept itself is still evolving: in contrast to its part related to cooperative interactions, of which a good understanding has already been obtained, several important problems raised in this concept had to date not been resolved satisfactorily or had not been studied extensively. After a careful, further examination and subsequent experimental evaluation of the CA action concept, we realized that the following, non-trivial issues needed attention and thorough investigation.

♦ Unlike transactions, a CA action can involve several cooperative activities (e.g. execution threads) which are jointly responsible for access to shared objects. The problem is how consistency of the shared objects can be still maintained, especially in the presence of concurrent access, failures and nested invocations.

♦ The derived objects and CA actions model extends the well-known objects and transactions model by providing additional support for cooperative concurrency. One major concern is how best to integrate and combine different kinds of concurrency within a unified framework while still keeping the overall system complexity well-controlled.

♦ So far as implementation issues be concerned, what is the possible structure of a CA action processing system in a distributed computing environment?

Analysis of, and discussion of possible solutions to, these problems are the subject of this paper. We introduce basic concepts, assumptions and definitions in section two. In section three, generalized transaction models are examined, and problems and difficulties with these models are discussed. We address the above three major issues in section four, together with proposed solutions. Finally, we present our conclusions in section five.

## 2: Fundamentals

### 2.1: Objects, Threads and Actions

In order to discuss the basic principles of our proposed framework, we must first introduce a simple model to describe the software systems we are considering. We define a *system* as a set of interacting objects. An *object* is a named entity that combines a data structure (internal state) with its associated operations; these operations determine the externally visible behaviour of the object. A *thread* is an active entity that is responsible for executing a sequence of operations on objects. Threads are the agents of computation. (Threads can exist syntactically, e.g. as in the Java language, or as a purely run-time concept.) A system is said to be *concurrent* if it contains multiple threads that behave as though they are all in progress at one time. In a *distributed* computing environment, this may literally true — several threads may execute at once, each on its own processing node.

In general, the *action* concept is a control abstraction that allows the application programmer to group a set of operations on objects into a logical execution unit. An action may be associated with some desirable properties. During the execution of the action, a variety of commit protocols are required to enforce corresponding properties. For example, atomic (trans)actions have the properties of 1) failure atomicity, 2) serializability and 3) permanence of effect and can be used to ensure consistency of shared objects even in the presence of failures and concurrent access. CA actions further emphasize the enclosure of multi-threaded cooperation and the strict prohibition of information smuggling into or out of the action boundaries [Xu *et al.* 1995] and hence facilitate coordinated error recovery. More precisely, a CA action must provide a recovery line (when relying on backward error recovery) in which the recovery points of the participating threads in the action are properly coordinated so as to avoid the *domino effect* [Randell 1975]. It must also provide a test line consisting of a set of acceptance tests. All the objects accessed by a CA action, may be shared with other actions, must invoke appropriate forward and/or backward recovery measures cooperatively once an error is detected inside the action, in order to reach some mutually consistent conclusion.

### 2.2: Kinds of Concurrency

There are at least three kinds of inter-thread concurrency [Hoare 1976]. *Independent* concurrency means concurrent threads have access to only disjoint object sets, without any form of sharing or interacting. *Competitive* concurrency implies that concurrent threads compete for some common objects, but without explicit cooperation. *Cooperative* concurrency occurs in many actual systems, e.g. real-time control applications, where concurrent threads cooperate and interact with each other in pursuit of some joint goal; each thread is responsible only for a part of the joint goal. Cooperation between concurrent threads may be based on various different forms of communication and interaction. We choose to model *inter-thread cooperation* as information transfer via shared objects. Such an abstraction may cover various actual forms of inter-thread cooperation, including inter-thread communication by updating shared objects that have some synchronization mechanisms, or by message passing (without requiring shared storage), and of inter-thread synchronization such as condition synchronization (usually no data passed) and exclusion synchronization (usually for shared object schemes).

### 2.3: Failure Assumptions

In this paper we assume that the hardware components of a distributed system are ordinary computers (or nodes) connected by a communication network. The objects that run on network nodes communicate with each other by message passing. In common with the assumptions made in most studies on fault-tolerant distributed systems (e.g. [Shrivastava et al. 1991]), a node will be assumed to work as specified or fail by stopping completely (crash). A crashed node may be repaired and made active again. A node can contain both stable storage and volatile storage. All data stored on volatile storage are assumed to be lost when a crash occurs; but data on stable storage are assumed to remain unaffected.

It is very important to notice that, in most models for fault-tolerant distributed systems, programs (or actions) that operate on objects are usually assumed to be correct (i.e. software-fault-free). However, residual software design faults in the code of an action could be a cause of some erroneous, but committed results. In fact, the CA action scheme [Xu et al. 1995] contains various integrated mechanisms for dealing with both hardware and software faults. In the interests of simplicity and for the sake of comparison with the transaction models, we concentrate in this paper on hardware-related faults only, i.e. if the system is known to be in an error-free state upon entry to an action, the action will be terminated normally (committed), producing the intended results, or aborted (due to hardware failures or conflict access), producing no results.

## 3: Transaction-Based Approaches for Supporting Cooperative Concurrency

### 3.1: Traditional Transactions

A flat (i.e. non-nested) transaction corresponds to a single execution thread composed of a partially ordered set of operations on objects and permits no concurrency inside the action. The nested transaction model [Moss 1981] extends the traditional transaction paradigm by providing the independent failure property for nested transactions and supports modular construction of applications. For error recovery purposes, nested transactions are in principle able to abort independently from their parent.

In these traditional models, programming in a (competitively) concurrent environment is made similar to that in a sequential environment. The application programmer does not need to worry about possible interference between concurrent and competitive actions — the system (i.e. those transparent mechanisms for concurrency control) will ensure interference-free access to shared objects despite concurrent invocations. However, this concurrency transparency leads to a serious limitation on the applicability of the transaction models: no appropriate and efficient support for multi-threaded cooperative activities [Wing et al. 1992].

### 3.2: Advanced Cooperative Transactions

There have been many generalizations of the basic transaction model, together representing the latest thinking in the area of transactions and databases on how to structure modern distributed and heterogeneous computations. We examined about 15 different proposals for advanced transaction models that attempt to overcome the limitations of the traditional transaction model (e.g. [Skarra 1989][Elmagarmid 1993][Taylor et al. 1994]), most of which provide to some extent support for cooperative activities.

Generally, these models are based on the concept of nested transactions. At the top-level, a generalized transaction has all the properties of traditional transactions, that is, failure atomicity, serializability, and permanence of effect. However, concurrent nested transactions may be allowed somehow to cooperate. With respect to the permitted degrees of cooperation between nested transactions, there are at least three major methods for extending traditional transactions. The first is to relax strict isolation [Taylor et al. 1994]: concurrent nested transactions must be still serializable but uncommitted results may be shared among them. A nested transaction that uses uncommitted data will depend on the nested transaction that produced the data. Such a nested transaction cannot commit or abort independently and may, once terminated, be required to wait for the commitment of any nested transaction on which it depends before committing. Since the serializability property must be satisfied, only limited cooperation between those nested transactions is possible.

The second approach is to enforce the user-defined execution order of nested transactions, which may be specified in the specification of the top-level transaction. The atomicity property may be kept but the mechanism for ensuring the correctness conditions defined by serializability must be extended to permit strong conditions defined by the user-specified execution order. Such cooperation is still restricted.

By combining the above two methods, a greater degree of cooperation between nested transactions could be achieved. Nodine and Zdonik [Nodine & Zdonik 1984] proposed the substitution of a notion of user-defined correctness for the notion of correctness defined by serializability. Because isolation is not required, correctness conditions on the execution order of operations involved in different (but cooperative) nested transactions could be defined for special application purposes. For example, the application programmer can define various correctness conditions based on the relationships like *conflicts*, *patterns* (i.e. specified order), and *triggers* etc [Skarra 1989]. Table 1 compares some common properties of generalized transactions with those of the basic nested transactions.

|  | *Generalized* Transactions | *Nested* Transactions |
|---|---|---|
| *Action on* | A set of objects | A set of objects |
| *Inter-concurrency* | Competitive sharing with other transactions | Competitive sharing with other transactions |
| *Intra-concurrency* | Cooperating subtransactions | Competing subtransactions |
| *Composed of* | Multiple concurrent but cooperating subtransactions, operations involved in these subtransactions can be specially ordered | Multiple concurrent and competing subtransactions, each composed of a sequence of partially ordered operations on the set of objects |
| *Correctness* | User-defined, no isolation | Serializability |

**Table 1** Comparison of Nested Transactions and Generalized Transactions.

### 3.3: Problems and Difficulties

Generalized transaction models start from the concept of traditional transactions and suffer inevitably from some original limitations of the traditional model. These models lose the atomicity property of nested transactions, leading to inconsistency between the top-level *atomic* transactions and nested transactions, and could therefore offer ambiguous semantics of a transaction. Moreover, loss of the atomicity property will complicate built-in recovery mechanisms and have negative impact on the system performance. On the other hand, cooperation among nested transactions, supported by these models, are still restricted since none of them permits true concurrent programming — the boundaries of nested transactions can only be opened up to a limited extent and explicit communications across the boundaries cannot be allowed (or otherwise semantic contradiction on the transaction notion could be caused). In the best case, the application programmer can use the specification mechanism provided by a model to define some operation conditions for related nested transactions and the system then in effect carries out some cooperation by enforcing the specified conditions. However, because no integrated mechanism for possible communications between nested transactions can be provided to the application programmer, it becomes a difficult task to achieve close and fine cooperation.

In contrast, the CA action concept is based on a totally different philosophy and has a much (system-)wider view on concurrent activities. A CA action allows concurrent threads to cooperate in performing a joint task. Explicit communication and coordination among threads are permitted completely but must be enclosed within the boundaries of a CA action. By the use of CA actions, most of the previously-identified limitations in generalized transaction models can be effectively overcome.

### 4: The Coordinated Atomic Action Approach

As described in [Xu *et al.* 1995], a CA action encloses a joint activity between a group of (two or more) interacting threads. Within the CA action, thread interactions could take place. How threads cooperate by means of general communication and synchronization facilities is well understood, so this issue will not discussed further. Rather, we will focus on how a CA action controls operations on common objects which may be shared with other actions and on what the fundamental distinctions are between CA actions and transactions.

### 4.1: Characteristics of CA Actions

A single (top-level) transaction is normally invoked by a single execution thread and consists of a partially ordered set of operations on a set of objects (see the example of multi-threaded transactions explained in [Wing *et al.* 1992]). In contrast, a CA action can be issued cooperatively by multiple execution threads. As the part of the joint computation, the CA action may contain a partially ordered set of operations on shared objects, *but* these operations can be invoked by different threads. Some operations (e.g. two operations on the same object) must be carried out in the order specified in the CA action, and the others (e.g. unrelated operations on different objects) may be carried out in any order or in parallel. Without considering any nesting of actions, we need at least a simple mapping system that translates those operation invocations from different threads into an operation sequence (in effect like an ordinary transaction) which the transaction mechanism integrated with CA actions can identify. By this means, the shared objects see no difference between a joint action and an independent transaction. This important property could also simplify implementation-related issues, given a suitable transaction processing system exists and can be used (see section 4.3).

It is extremely important to notice that the execution order of operations invoked in a CA action is determined by both the order specified by each participating thread *and* the communication and synchronization relationship between participating threads. Unfortunately, any nesting of CA actions makes the task of developing a *simple* mapping mechanism very difficult. In general, nested transactions allow intra-transaction concurrency between nested transactions. The concurrency control algorithm in [Moss 1981] permits concurrent access to shared objects from the nested transactions at the same level of nesting, based on the serializability property. However, the nesting relationship in CA actions has more complex semantics: a nested CA action may contain just a subset of the participating threads of the parent action and hence can occur concurrently with other participating threads. Moreover, two nested CA actions at the same level of nesting may have to be executed according to some specified order. In any case, the correctness conditions defined by serializability are no longer sufficient for the operation invocations, which instead must adhere to some application-specific ordering conditions. It is obvious that there is no one-to-one mapping between the nesting relationship of a given set of CA actions and that of ordinary transactions. For example, a CA action that has only one nested action might well correspond to a transaction that contains one, or two or multiple nested transactions, much depending upon the application-specific cooperation relationship between the participating threads. Figure 3 shows some details of the comparison of nested CA actions and nested transactions.

| | A *top-level* CA action | A *top-level* transaction |
|---|---|---|
| *Action on* | A set of objects | A set of objects |
| *Inter-concurrency* | Competitive sharing with other actions | Competitive sharing with other actions |
| *Intra-concurrency* | Cooperating threads **and** nested CA actions | Competing nested transactions (multiple sub-threads) |
| *Composed of* | A set of partially ordered operations, invoked by cooperating threads **and** sub-actions | Multiple concurrent nested transactions, each composed of a set of partially ordered operations on the set of objects |
| *Participants* | Multi-threads, but one as the trivial case | One thread for the top-level transaction and multiple sub-threads for nested transactions |

**Table 2** Comparison of CA actions and Transactions.

There are two ways of structuring the (non one-to-one) mapping mechanism. The first is to translate the set of operations in nested CA actions into a flat transaction. This approach can keep the translation algorithm simple but may lose a great degree of concurrency. A more complicated scheme is to translate the operation set of nested CA actions into those of the equivalent nested transactions. The specially ordered part in the nested CA actions may be translated into sequential nested transactions while the concurrent and unrelated part may well correspond to concurrent (serializable) nested transactions.

We are aware that a new mechanism that directly controls operations on shared objects, designed specially for CA actions, would allow a greater degree of concurrency, and achieve better performance, than a transaction mechanism plus the corresponding mapping mechanism (for example see [Wing *et al.* 1992]). However, given the widespread existence of various transaction processing systems and mechanisms, an implementation derived from such existing mechanisms might be more realistic from the practical point of view.

In comparison with generalized transaction models for handling cooperative activities, the CA action model has many favourable characteristics. First, the atomicity property holds at any level of nesting. Secondly, since the application programmer is responsible for designing cooperation and coordination, at least in principle, such application-specific cooperation will potentially permit a greater degree of concurrency. Thirdly, the application programmers have full flexibility in defining interactions and coordination between concurrent threads by ordinary language mechanisms for concurrent programming, but they do so in a strictly disciplined manner.

### 4.2: Objects and CA Actions Model: Supporting different kinds of concurrency

We present the objects and CA actions model here derived from the CA action concept. We will not consider again several important parts of the derived model, such as basic concepts, failure assumptions, and definitions related to CA actions, which have been examined above, and in [Xu *et al.* 1995], but will concentrate on how best to combine both cooperative and competitive concurrency within the model in a natural and controlled way. With respect to different degrees of implementation difficulties, integration of various kinds of (e.g. independent, competitive, or cooperative) concurrency may take different forms — from simple to complex.

*The flat model:* Without nesting, for a given system, the concurrency relationships between concurrent execution threads of the system can be clearly defined. The threads that are not in any action must execute independently without having impact on any other threads and actions. Single-threaded actions behave like ordinary transactions and are serializable with any other actions (i.e. competitive concurrency). Within a multi-threaded CA action (i.e. one that contains two or more participating threads) cooperative concurrency is permitted. There are several trivial cases: the system may contain just independent threads, or competitive actions, or a global CA action that has all the execution threads as its participants.

*The nested action model:* This model allows only the nesting of the same kinds of concurrency, that is, a single-threaded, transaction-like, action permits just competitive concurrency among nested actions while a multi-threaded CA action provides support for cooperative concurrency only. At the top level, various CA actions and independent threads can co-exist, exactly like the flat model above. At the further levels of nesting, a single-threaded CA action can contain nested actions which are all single-threaded so that the usual nested transaction mechanisms can be used to ensure their correctness; but a multi-threaded CA action may have nested CA actions which, in turn, allow for only cooperative concurrency in the way that normal coordination mechanisms may be used to ensure application-specific correctness conditions.

*The general model* could however permit the complex nesting of different kinds of concurrency, supporting a great degree of concurrency and high flexibility. For example, a CA action that encloses cooperative activities might have some nested actions which can behave like nested transactions, that is, their child actions do not cooperate but compete for shared objects. On the other hand, a single-threaded CA action might have nested CA actions which clearly enclose cooperative concurrency. This model certainly causes new implementation difficulties and the control mechanisms that switch over different kinds of concurrency must be provided.

### 4.3: Structure of a CA Action Processing System

Figure 1 outlines a possible structure of distributed transaction processing systems (see a more detailed description in [Bacon 1993]). A client submits a transaction at one node to the processing system. Note that the system may work on several submissions concurrently. An object resides at one and only one node, and hence some mechanisms for locating an object are required, given its unique identifier. The transaction manager is responsible for validating the clients' submissions, for locating objects invoked by the transaction, and for passing the operations contained by the transaction to the (local or remote) scheduler. The scheduler will use some strategy to achieve a serializable schedule of the operations of the transactions in progress.

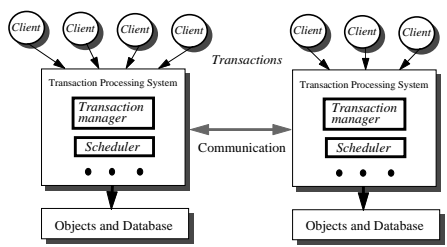

**Figure 1** A Distributed Transaction Processing System.

The transaction manager at a single node receives a client request for a transaction and initiates local and remote operation invocations. It must be notified of the results of attempted invocations: whether an invocation was accepted and completed, or rejected, or perhaps aborted for some reason. Assuming that all the transaction's invocations at all the nodes have been completed, the transaction must then be committed collectively. Other integrated mechanisms must be also provided including those for controlling error recovery and managing object states for both long-term and short-term storage.

It is however obvious that such an implementation will not work for the more complex case of CA actions. A CA action may be submitted jointly from different clients who can make their contributions to this submission from different nodes. Moreover, a CA action manager is responsible for many additional, new management jobs. To permit CA actions, two new integrated mechanisms must be provided, one to coordinate multiple participating threads and one to map a CA action (or nested CA actions) to an ordinary transaction (or to nested transactions) which could be identified and managed by the underlying transaction processing system.

The basic functions of the CA action manager is 1) to register asynchronous entries, perhaps in different nodes, of the participating threads for a given action; 2) to enforce the correct nesting of nested CA actions; and 3) to synchronize the exit of all participants. A more detailed discussion can be found in [Xu *et al.* 1995]. The functions of the mapping sub-systems are to translate CA actions into transactions whenever consistent access to shared objects is needed. Two possible algorithms were addressed in section 4.1 and may be well incorporated into the structure illustrated in Figure 2.

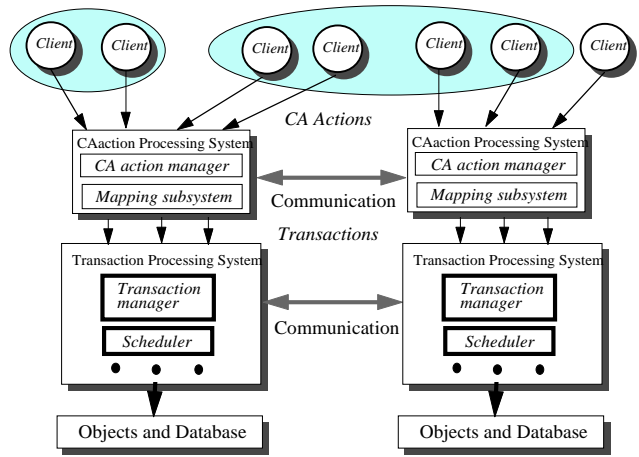

**Figure 2** A Distributed CA Action Processing System.

### 4.4: A Prototype Implementation

Our implementation was organized in a distributed environment that consists of a set of SUN workstations (running UNIX) connected through a communication network (TCP/IP). We used Java as the major programming language in order to establish an architectural connection to an existing fault-tolerant distributed system — Arjuna [Shrivastava *et al.* 1991][Parrington *et al.* 1995] — built at Newcastle. The Arjuna system was originally developed in C++ and currently re-implemented in Java based on the objects and (trans)actions model without direct support for cooperative concurrency. Our implementation adds a new API-like layer on the top of the Arjuna system and provides full support for the objects and CA actions model. The delicate problems tackled and results obtained from our case studies and prototypes are however beyond the intended scope of this paper and can be found in [Zorzo et al. 1997][Xu et al. 1998].

## 5: Concluding Remarks

Real-world applications contain many examples of cooperative concurrent activities, but the widely-used objects and (trans)actions model, in fact, does not provide appropriate support for cooperative concurrency. Generalized transaction schemes provide only limited support for cooperation by somehow opening up nested transactions. Because cooperative activities are application-specific in nature, a general mechanism for total concurrency transparency cannot be obtained. Nevertheless, coordinated atomic actions attempt to enclose multi-execution threads whose interactions and cooperation can be explicitly specified by application programmers, and hence support the greatest degree of cooperative concurrency while still maintaining the atomicity property for all actions.

By contrast with the objects and (trans)actions model, our objects and CA actions model can be used to construct fault-tolerant distributed systems that provide full support for both cooperative and competitive concurrency. With respect to different degrees of system complexity and implementation difficulties, this new model may well take different forms, from simple (e.g. flat actions) to complex (e.g. more general, nested actions with a variety of correctness properties). For flat (non-nested) CA actions, ordinary transaction mechanisms can be exploited for ensuring consistent access to shared objects. However, existing nested transaction mechanisms cannot be used directly to support nested cooperative activities; extra mechanisms that control multi-threads' rendezvous and map nesting relationships must be developed. To allow for full nesting of different kinds of concurrency, further additional mechanisms are needed for controlling correct (and automatic) switches and combination of various actions with specified correctness properties.

The objects and (trans)actions model assumes that the transaction itself is correct (software-fault-free). Such an assumption may be questionable, especially in a distributed and heterogeneous environment. The objects and CA actions model requires explicit concurrent programming, and so software design faults might be of greater concern. Associated mechanisms for program error detection, both forward and backward error recovery, and software redundancy merit further investigation and must be fitted into the CA action framework properly.

## Acknowledgements

This work was supported by the ESPRIT Basic Research Projects 3092 and 6362 on "Predictably Dependable Computing Systems", and by the ESPRIT Long Term Research Project 20072 on "Design for Validation" (DeVa).